\documentclass{iaus}
\usepackage{graphicx}

\title[Magellanic star formation] 
{Triggered star formation in the Magellanic Clouds}

\author[K. Bekki]   
{Kenji  Bekki$^1$}%

\affiliation{$^1$ School of Physics, University of New South Wales, 
Sydney 2052, Australia
\break email: bekki@phys.unsw.edu.au}

\pubyear{2006}
\volume{237}  
\pagerange{119--126}
\date{?? and in revised form ??}
\jname{Triggered Star Formation in a Turbulent ISM}
\editors{B. G. Elmegreen \& J. Palous, eds.}
\begin{document}

\maketitle

\begin{abstract}

We discuss  how tidal interaction 
between the Large Magellanic Cloud (LMC), the Small Magellanic Cloud (SMC),
and the Galaxy triggers galaxy-wide
star formation in the Clouds 
for the last $\sim 0.2$ Gyr based on our chemodynamical simulations 
on the Clouds.
Our simulations demonstrate that the tidal interaction
induces the formation of
asymmetric spiral arms with high gas densities
and consequently  triggers star formation within the arms in the LMC.
Star formation rate in the present LMC
is significantly enhanced just above the eastern edge of the LMC's
stellar bar owing to the tidal interaction.  
The location of the enhanced star formation is very similar to
the observed location of  30 Doradus, which suggests that
the formation of 30 Doradus is closely associated with
the last Magellanic collision about 0.2  Gyr ago.
The tidal interaction can dramatically compress
gas initially within the outer part of the SMC
so that new stars can be formed from the gas to become 
intergalactic young stars in the inter-Cloud region
(e.g., the Magellanic Bridge).
The metallicity distribution function of the newly formed stars
in the Magellanic Bridge
has a peak of [Fe/H] $\sim$ $-0.8$, which is significantly lower
than the stellar metallicity of the SMC.

\end{abstract}

\firstsection 
\section{Introduction}

The Magellanic system composed of the LMC and the SMC is 
believed to be an interacting one where
star formation histories of the Clouds
have been strongly influenced
by dynamical and hydrodynamical  effects of galaxy  interaction 
(\cite[Westerlund 1997]{Westerlund97}).
It is however unclear how galaxy interaction
between the Clouds and the Galaxy 
triggers star formation in the gas disks of the Clouds.
Recent observations on spatial distributions of HI 
(\cite[Staveley-Smith
\etal\ 2003]{Staveley-Smith03}),
molecular gas (\cite[Fukui
\etal\ 1999]{Fukui99}),
and young stars (\cite[Grebel \&  Brandner 1998]{Grebel98})
have provided vital information on   
galaxy-wide triggering mechanisms of star formation 
in the Clouds.
By comparing numerical simulations of the Magellanic system
with these observations,
we here discuss (1) how the tidal interaction changes
the spatial distribution of high-density gaseous regions
where new stars can be formed in the LMC, (2) whether the formation of
30 Doradus is triggered by the interaction,
and (3) how the interaction triggers star formation
in the Magellanic Bridge (MB). 


\begin{figure}
\centerline{
\scalebox{0.5}{%
\includegraphics{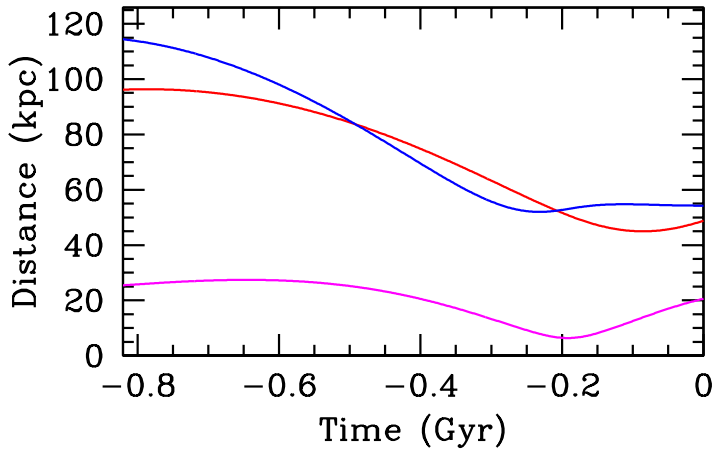}%
}
}
\caption{Time evolution of distance between
the LMC and the SMC (magenta), the LMC and the Galaxy (red),
and the SMC and the Galaxy (blue) for the last 0.8 Gyr.
Note that the LMC-SMC distance becomes minimum (8kpc) about
0.2 Gyr ago.}
\end{figure}

\section{The last Magellanic interaction}

We investigate the last 0.8 Gyr evolution of the Clouds
orbiting the Galaxy based on GRAPE chemodynamical simulations
of the Clouds
with star formation models (\cite[Bekki \etal\ 2004]{Bekki04};
\cite[Bekki \& Chiba 2005]{Bekki05};
\cite[Bekki \& Chiba 2006]{Bekki06}). 
Since the details of the numerical methods and the initial conditions
of the Clouds have been already discussed in our previous papers,
we here summarize the models briefly.
The total masses of the LMC and the SMC are set to be $2.0 \times 10^{10}
{\rm M}_{\odot}$ and $3.0 \times 10^{9} {\rm M}_{\odot}$,
respectively. Gas particles are assumed to be converted into new stars
according to the   Schmidt law with the observed threshold
gas density (\cite[Kennicutt 1998]{Kennicutt98}).
Figure 1 shows that the pericenter distance of
the SMC orbit with respect to the LMC is 8 kpc about 0.2 Gyr ago.
The tidal force from the LMC is therefore about 20 times stronger
than that from the Galaxy for the SMC, which means that the SMC
can be more strongly influenced by the LMC-SMC interaction than
the SMC-Galaxy one. This LMC-SMC interaction can also significantly
influence the gaseous evolution of the LMC and thus its recent 
star formation history.

\begin{figure}
\centerline{
\scalebox{0.8}{%
\includegraphics{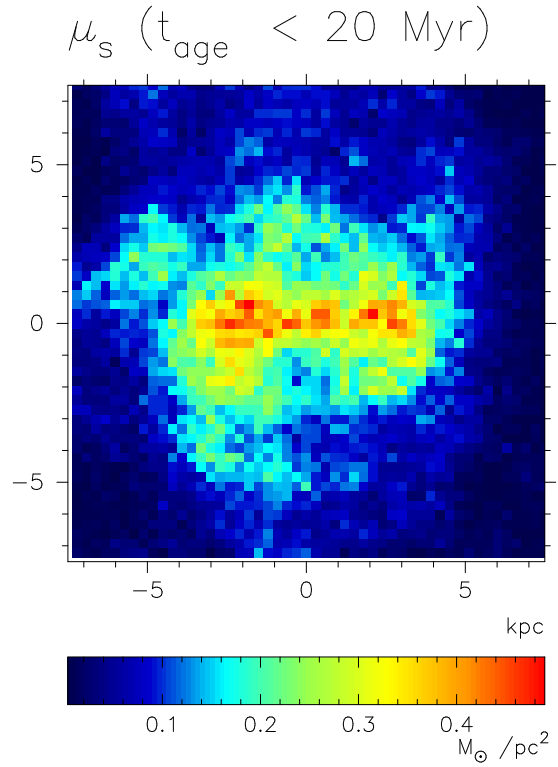}%
}
}
\caption{
The projected distribution of surface mass densities of young stars
with ages less than 20 Myr formed in the LMC during the
LMC-SMC-Galaxy interaction.
}
\end{figure}
\section{Formation of 30 Doradus}

Strong tidal effects of the LMC-Galaxy and the SMC-LMC interaction
can induce the formation of asymmetric spiral arms with high densities
of gas so that new stars can be formed in the arms.  Figure 2
demonstrates that (1) the spatial distribution of young stars
is quite irregular and clumpy, (2) there is a strong concentration
of young stars along the stellar bar (composed of old stars),
and (3) there is an interesting peak just above the eastern
edge of the bar.  This interesting peak of the stellar density
of very young stars corresponds to the location where two
asymmetric spiral arms emerge in the LMC disk.
The location of the peak is very similar to the location
of 30 Doradus, which suggests that the formation of 30 Doradus
is closely associated with the formation of strong spiral
arms due to the last Magellanic interaction about 0.2 Gyr ago.
The simulated two high-density gaseous arms
in eastern and southern parts of the LMC  are morphologically
similar to the observed gaseous arms composed of molecular
clouds in the southern part of the LMC (i.e., ``the molecular
ridge''). This similarity suggests that the origin of
the observed peculiar distributions of molecular clouds 
(\cite[Fukui
\etal\ 1999]{Fukui99})
is due to the recent Magellanic interaction.  
The mean star formation rate of the LMC is increased rapidly
by a factor of 5 about 0.2 Gyr ago and the rapid increase
is synchronized with
the enhancement of the star formation rate of the SMC
in our models.

\begin{figure}
\centerline{
\scalebox{0.4}{%
\includegraphics{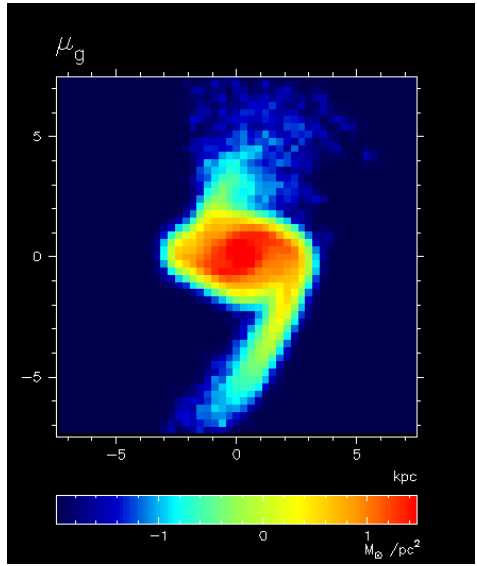}%
}
}
\caption{
The projected distribution of smoothed (column) gas densities of the SMC about 
0.14 Gyr ago (i.e., 60 Myr after the last Magellanic collision).
The lower tidal tail with a higher gas density is the forming MB. 
}
\end{figure}

\section{Star formation in the MB}

The tidal interaction can also significantly change the recent star formation
histories
not only in the central region of the SMC's gas disk  but also in
its outer part, which finally becomes the MB
after the interaction. Figure 3 shows that
the SMC's outer gas disk is strongly compressed by the interaction
so that gas densities along 
the forming MB can exceed the threshold
gas density of star formation (i.e., $3 {\rm M}_{\odot}$ pc$^{-2}$).
Since the MB is formed from the outer gas, where the metallicity
is significantly smaller owing to the negative metallicity gradient
of the SMC,  the metallicity distribution of new stars in the MB
shows a peak of [Fe/H] $\sim$ $-0.8$ (i.e., 0.2 dex smaller than
the central stellar metallicity of the simulated SMC).
About 25\% of the initial gas mass of the SMC is finally distributed
in the MB whereas only 0.1\% of the gas mass is converted into
new stars in the MB.
The present model thus provides a physical explanation for the
origin of the observed formation sites of new stars along
the MB 
(\cite[Mizuno  \etal\ 2006]{Mizuno06}).

\section{Conclusions}

The present study
has suggested that
tidal interaction between the Clouds and the Galaxy is 
closely associated with the formation of 30 Doradus,
the southern molecular ridge of the LMC,  and inter-Cloud stars with
low metallicities. The observed asymmetric and clumpy distributions of 
young stars in the LMC are  demonstrated
to be due to the last Magellanic interaction about 0.2 Gyr ago,
which forms asymmetric spiral arms with high-density gas.
The synchronized burst of star formation
in the Clouds about 0.2 Gyr ago  
(by tidal interaction)
can be proved by
the observed age distributions of young star clusters in the Clouds
(e.g., \cite[Girardi
\etal\ 1995]{Girardi95}).
The simulated distributions of star-forming regions 
will be compared with the latest results of the Spitzer observations
on young stellar objects (YSOs) in the Clouds.



\end{document}